\documentclass[12pt,A4]{article}
\usepackage{amssymb,amsfonts,amsthm,amsmath}

\usepackage{epsf}
\usepackage{graphicx}
\usepackage{placeins}
\usepackage{setspace}

\def\hang{\hangindent\parindent}
 \def\rf{\par\noindent\hang}

\begin{document}
\baselineskip=21pt

\begin{center} \Large{{\bf
Computation of confidence intervals in regression utilizing uncertain prior information}}
\end{center}

\bigskip

\begin{center}
\large{
{\bf Paul Kabaila$^*$, Khageswor Giri}}
\end{center}

\begin{center}
{\sl Department of Mathematics and Statistics,
La Trobe University, Victoria 3086, Australia}
\end{center}

\medskip

\noindent {\bf ABSTRACT}

\medskip

\noindent We consider a linear regression model with regression parameter $\beta = (\beta_1, \ldots,  \beta_p)$ and
independent and identically $N(0, \sigma^2)$ distributed errors.
Suppose that the parameter of interest is $\theta = a^T \beta$ where $a$ is a specified vector.
Define the parameter $\tau = c^T \beta - t$ where the vector $c$ and the number $t$
are specified and $a$ and $c$ are linearly independent.
Also suppose that we have uncertain prior information that $\tau = 0$.
Kabaila and Giri (2009c) present a new frequentist $1-\alpha$ confidence interval for $\theta$ that utilizes this
prior information. This interval has expected length that (a) is relatively small when the prior
information about $\tau$ is correct and (b) has a
maximum value that is not too large. It coincides with the standard $1-\alpha$ confidence interval
(obtained by fitting the full model to the data) when the data strongly contradicts the prior information.
At first sight, the computation of this new confidence interval seems to be infeasible.
However, by the use of the various computational devices that are presented in detail in the present paper,
this computation becomes feasible and practicable.

\bigskip


\vspace{0.7cm}

\noindent $^*$Corresponding author. Tel.: +61 3 9479 2594, fax: +61 3 9479 2466.

\noindent {\sl E-mail address:} P.Kabaila@latrobe.edu.au (Paul Kabaila).

\newpage

\noindent {\bf 1. Introduction}

\medskip

 Consider the linear regression model
 \begin{equation*}
 Y = X \beta + \varepsilon
 \end{equation*}
 where $Y$ is a random $n$-vector of responses, $X$ is a known $n \times p$ matrix with linearly
independent columns, $\beta = (\beta_1,\ldots, \beta_p)$ is an unknown parameter vector and
$\varepsilon \sim N(0, \sigma^2 I_n)$ where $\sigma^2$ is an unknown positive parameter.
Suppose that the parameter of interest is $\theta = a^T \beta$ where $a$ is specified
$p$-vector ($a \ne 0$). Define the parameter $\tau = c^T \beta - t$ where the vector $c$ and the number $t$
are specified and $a$ and $c$ are linearly independent.
Also suppose that previous experience with similar data sets and/or
expert opinion and scientific background suggest that $\tau = 0$.
In other words, suppose that we have uncertain prior information that $\tau = 0$.
Examples include having uncertain prior information that
(a) one of the regression coefficients $\beta_i$ takes a specified value and
(b) the linear regression consists of two parallel straight line regressions.
``Higher order'' terms in a linear regression model are often strong candidates for
terms that could plausibly be zero. For example, for factorial experiments it is
commonly believed that three-factor and higher order interactions are negligible. Indeed, this type of
belief is the basis
for the design of fractional factorial experiments. Another example is that it is
commonly believed that the highest order terms in a univariate or multivariate polynomial regression
are likely to be negligible.
Our aim is to find a frequentist $1-\alpha$ confidence interval (i.e. a confidence
interval whose coverage probability has infimum $1-\alpha$)
for $\theta$ that utilizes this uncertain
prior information, based on an observation of $Y$.

One may attempt to utilize the uncertain prior information as follows. We carry out a preliminary test
of the null hypothesis $\tau = 0$, against the alternative hypothesis $\tau \ne 0$.
We then find the confidence interval for $\theta$, with nominal coverage $1-\alpha$,
based on the assumption that the selected model had been given to us {\it a priori}.
It might be hoped that this confidence interval will have good coverage properties and an expected length that (a) is relatively
small when the prior information is correct and (b) is not too large when the prior information happens
to be incorrect.
This assumption is false and, as pointed out by Kabaila (1995, 1998, 2005, 2009), Giri and Kabaila (2008),
Kabaila and Giri (2009b) and Kabaila
and Leeb (2006), it leads to a confidence interval whose minimum coverage is typically
far below $1-\alpha$. In other words, this confidence interval fails abysmally to utilize the uncertain
prior information.

We assess a $1-\alpha$ confidence interval for $\theta$ using the ratio
(expected length of this confidence interval)/(expected length of standard $1-\alpha$ confidence interval).
The standard $1-\alpha$ confidence interval is obtained by fitting the full model
to the data.
We call this ratio the scaled expected length of this confidence interval.
Kabaila and Giri (2009c)
describe a new $1-\alpha$ confidence interval for $\theta$ that utilizes the
prior information. This interval has scaled expected length that (a) is
substantially smaller than 1 when the prior
information that $\tau=0$ is correct and (b) has a
maximum value that is not too much larger than 1.
It coincides with the standard $1-\alpha$ confidence interval
when the data strongly contradicts the prior information.
This interval also has the attractive property that it has endpoints that are
continuous functions of the data.

Let $\hat \Theta$ and $\hat \tau$ denote the least squares estimator of $\theta$
and $\tau$ respectively. Define the correlation coefficient
$\rho = \text{corr}(\hat \Theta,\hat \tau)$.
Also define the parameter $\gamma = \tau/\sqrt{\text{var}(\hat \tau)}$. Both the
coverage probability and the scaled expected length of the new $1-\alpha$ confidence interval
are even functions of $\gamma$. An example of the performance of this confidence interval
is shown in Figure 2 for the case that $1-\alpha = 0.95$, $n-p=1$ and $\rho=0.4$.
The top panel of this figure is a plot of the coverage probability of the new 0.95 confidence
interval for $\theta$ as a function of $\gamma$. This plot shows that this coverage
probability is 0.95 throughout the parameter space. The bottom panel of Figure 2 is a plot
of the square of the scaled expected length of this confidence interval as a function of
$\gamma$. When the prior information is correct (i.e. $\gamma=0$), we gain since the square of the scaled expected length
is substantially smaller than 1. The maximum value of the square of the scaled expected length is not too large.
The new 0.95 confidence interval for $\theta$ coincides with the standard $1-\alpha$ confidence interval
when the data strongly contradicts the prior information. This is reflected in Figure 2 by the fact that the
square of the scaled expected length approaches 1 as $\gamma \rightarrow \infty$.
All computations presented in the paper were performed with programs written in MATLAB, using the Optimization
and Statistics toolboxes.

In Section 2, we describe the constrained minimization problem that needs to be solved to find
this new confidence interval. To arrive at this description, Kabaila and Giri (2009c) have already
used the following simplification techniques: (a) invariance arguments that take account of the form
of the uncertain prior information, (b) a simply-implemented constraint on the new confidence interval
that guarantees that it will coincide with the standard $1-\alpha$ confidence interval
when the data strongly contradicts the prior information and (c) simplified
expressions for the coverage probability and the criterion to be minimized. Even so,
the coverage probability constraint portion of this minimization problem involves
a continuum of constraints. Thus, at first sight, the computation of this new confidence interval
seems to be infeasible. In Section 3 we describe how this continuum of constraints can be
replaced by a finite number or appropriately-chosen constraints. Even though this makes the
computation of the new confidence interval feasible, a significant number of computational issues
remain to be solved. The solution to these computational issues is
described by Giri (2008) and presented in detail
in the present paper. We compute the double integrals for the
coverage probability, scaled expected length and the criterion to be minimized by first truncating these integrals. In Section
4, we present bounds on the resulting truncation errors. In Section 5 we present
some practical advice on how to make these computations work. In Section 6, we present a numerical example that illustrates
the successful computation of the new confidence interval.

\bigskip

\noindent {\bf 2. Constrained minimization problem to be solved}

\medskip

Let $\hat \beta$ denote the least squares estimator of $\beta$. Let
$\hat \Theta$ denote $a^T \hat \beta$ i.e. the least squares estimator of $\theta$.
Also, let $\hat \tau$ denote $c^T \hat \beta - t$ \ i.e. the least squares estimator of $\tau$.
Define the matrix $V$ to be the covariance matrix of $(\hat \Theta, \hat \tau)$ divided by $\sigma^2$.
Let $v_{ij}$ denote the $(i,j)$ th element of $V$.
The standard $1-\alpha$ confidence interval for $\theta$
(obtained by fitting the full model to the data) is
$I = \big [ \hat \Theta - t_{n-p,1-\frac{\alpha}{2}} \sqrt{v_{11}} \hat
\sigma, \quad \hat \Theta + t_{n-p,1-\frac{\alpha}{2}} \sqrt{v_{11}} \hat
\sigma \big ]$,
where the quantile $t_{m,a}$ is defined by $P(T \le t_{m,a}) = a$ for $T \sim t_m$ and
$\hat \sigma^2 = (Y - X \hat \beta)^T (Y - X \hat \beta)/(n-p)$.

We use the notation $[a \pm b]$ for the interval $[a-b, a+b]$ ($b > 0$).
Define the following confidence interval
for $\theta$
\begin{align*}
\label{J(b,s)}
J(b,  s) = \bigg [ \hat \Theta -
\sqrt{v_{11}} \hat \sigma \, b\bigg(\frac{\hat{\tau}}{\hat \sigma \sqrt{v_{22}}}\bigg) \, \pm \,
\sqrt{v_{11}} \hat \sigma \, s\bigg(\frac{|\hat{\tau}|}{\hat \sigma \sqrt{v_{22}}}\bigg)
\bigg ]
\end{align*}
where the functions $b$ and $s$ are required to satisfy the following restriction.
\smallskip

\noindent {\underbar{Restriction 1}}\ \ \ $b: \mathbb{R} \rightarrow \mathbb{R}$  is
an odd function and $s: [0, \infty)
\rightarrow [0, \infty)$.

\smallskip
\noindent The motivation for this restriction is provided by the invariance arguments presented in Appendix A
of Kabaila and Giri (2009c).
We also require that the functions $b$ and $s$
satisfy the following restriction.
\smallskip

\noindent {\underbar{Restriction 2}} \ \ \
$b$ and $s$ are continuous functions.

\smallskip
\noindent This implies that the endpoints of the confidence interval $J(b,s)$ are continuous functions
of the data. Finally, we require the confidence interval $J(b,s)$ to coincide with
the standard $1-\alpha$ confidence interval $I$ when the data strongly contradict
the prior information. The statistic $|\hat \tau|/(\hat \sigma \sqrt{v_{22}})$ provides some indication
of how far away $\tau / (\sigma \sqrt{v_{22}})$ is from 0.
We therefore require that
the functions $b$ and $s$
satisfy the following restriction.
\smallskip

\noindent {\underbar{Restriction 3}} \ \ \
$b(x)=0$ for all $|x| \ge d$ and $s(x)=t_{n-p,1-\frac{\alpha}{2}}$ for all $x \ge d$
where $d$ is a (sufficiently large) specified positive number.

\medskip

Define $\rho = \text{corr}(\hat \Theta,\hat \tau)
= v_{12}/\sqrt{v_{11} v_{22}}$,
$\gamma = \tau/\sqrt{\text{var}(\hat \tau)} = \tau/(\sigma \sqrt{v_{22}})$
and $W = \hat \sigma/\sigma$.
Let $m = n-p$.
Note that $W$ has the same distribution as $\sqrt{Q/m}$ where $Q
\sim \chi^2_m$.
Let $f_W$ denote the probability density function of $W$.
Note that $f_W(w) = 2 m w f_m(m w^2)$ for all $w > 0$, where
$f_m$ denotes the $\chi^2_m$ probability density function.

For given $b$, $s$ and $\rho$, the coverage probability $P\big( \theta \in J(b, s) \big)$ is a function of $\gamma$.
We denote this coverage probability by $c(\gamma;b,s, \rho)$.
Part of our evaluation of the confidence interval $J(b,s)$ consists of comparing it with the
standard $1-\alpha$ confidence interval
$I$ using the criterion (expected length of $J(b, s)$)/(expected length of $I$).
We call this the scaled expected length of $J(b, s)$.
This is an even function of $\gamma$, for given $s$. We denote this function by $e(\gamma;s)$.

Our aim is to find functions $b$ and $s$ that satisfy Restrictions 1--3 and such that (a) the infimum of $c(\gamma;b,s, \rho)$ over $\gamma$ is
$1-\alpha$ and (b)
\begin{equation}
\label{criterion}
\int_{-\infty}^{\infty} (e(\gamma;s) - 1) \, d \nu(\gamma)
\end{equation}
is minimized, where the weight function $\nu$ has been chosen to be
\begin{equation}
\label{mixed_wt_fn}
\nu(x) = \lambda x + {\cal H}(x) \ \text{ for all } \ x \in \mathbb{R},
\end{equation}
where $\lambda$ is a specified nonnegative number and
${\cal H}$ is the unit step function defined by ${\cal H}(x) = 0$ for $x < 0$ and
${\cal H}(x) = 1$ for $x \ge 0$.
This weight function has also been used by Farchione and Kabaila (2008) and
Kabaila and Giri (2009a).
The larger the value of $\lambda$, the smaller the relative weight given to
minimizing $e(\gamma;s)$ for $\gamma=0$, as opposed to
minimizing $e(\gamma;s)$ for other values of $\gamma$.
For appropriately chosen $\lambda$, the weight function \eqref{mixed_wt_fn} leads to a $1-\alpha$
confidence interval for $\theta$ that has expected length that (a) is relatively small
when $\tau=0$ and (b) has maximum value that is not too large.

The following theorem provides computationally convenient expressions
for the coverage probability and scaled expected length of $J(b, s)$.

\medskip

\noindent {\bf Theorem 1.} (Kabaila and Giri (2009c)).
{\sl

\noindent (a) Define $\ell(h,w) = b(h/w)\, w - s(|h|/w) \, w$  and
$u(h,w) =  b(h/w)\, w + s(|h|/w) \, w$. Also define $\Psi(x, y; \mu, v) = P(x \le Z \le y)$ for $Z \sim N(\mu,v)$.
Now define the functions
$k(h,w,\gamma, \rho) = \Psi \left(\ell(h,w), u(h,w); \rho(h-\gamma),1-\rho^2 \right )$
 and
$k^{\dag}(h,w, \gamma, \rho) = \Psi \left( -t_{n-p,1-\frac{\alpha}{2}} w, t_{n-p,1-\frac{\alpha}{2}} w;
 \rho(h-\gamma), 1-\rho^2 \right )$.
The coverage probability of $J(b, s)$ is equal to
\begin{equation}
\label{cov_prob}
(1-\alpha) +
\int_0^{\infty} \int_{-d}^{d} \big( k(wx,w, \gamma, \rho) - k^{\dag}(wx,w, \gamma, \rho) \big)
\, \phi(wx-\gamma)\,  dx \, w \, f_W(w) \, dw
\end{equation}
where $\phi$ denotes the $N(0,1)$ probability density function.
For given $b$, $s$ and $\rho$, $c(\gamma;b,s, \rho)$ is an even function of $\gamma$.

\smallskip

\noindent (b) The scaled expected length of $J(b, s)$
is denoted $e(\gamma;s)$ and is equal to
\begin{equation}
\label{comp_conv_e}
1 +
 \frac{1} {t_{n-p, 1 - \frac{\alpha}{2}} \, E(W)}
 \int^{\infty}_0 \int^{d}_{- d} \left (s(|x|) - t_{n-p,1-\frac{\alpha}{2}} \right )
\phi(w x -\gamma) \, dx \,  w^2 \,  f_W(w)  \, dw.
\end{equation}
}

\medskip

The method used to compute $E(W)$ is described in Appendix A.
Substituting \eqref{comp_conv_e} into \eqref{criterion}, we obtain that \eqref{criterion} is equal to
\begin{equation*}
 \frac{2} {t_{n-p, 1 - \frac{\alpha}{2}} \, E(W)}
 \int^{\infty}_0 \int^{d}_{0} \left (s(x) - t_{n-p,1-\frac{\alpha}{2}} \right )
(\lambda + \phi(w x)) \, dx \,  w^2 \,  f_W(w)  \, dw
\end{equation*}
This is proportional to
\begin{align}
\label{criterion_final}
 &\int^{\infty}_0 \int^{d}_{0} \left (s(x) - t_{n-p,1-\frac{\alpha}{2}} \right )
(\lambda + \phi(w x)) \, dx \,  w^2 \,  f_W(w)  \, dw  \notag \\
&=\lambda \left (\int_0^d s(x) \, dx - d \, t_{n-p,1-\frac{\alpha}{2}} \right ) +
\int^{\infty}_0 \int^{d}_{0} \left (s(x) - t_{n-p,1-\frac{\alpha}{2}} \right )
\phi(w x) \, dx \,  w^2 \,  f_W(w)  \, dw,
\end{align}
since $E(W^2)=1$.
Therefore, our aim is to find functions $b$ and $s$ that satisfy Restrictions 1--3
and such that \eqref{criterion_final}
is minimized with respect to the functions $b$ and $s$, subject to the constraint that
\eqref{cov_prob} $\ge 1 - \alpha$
for all $\gamma \ge 0$.

Unless we specify parametric forms for the
functions $b$ and $s$, the computation of these functions
(to solve the constrained minimization problem) will certainly be infeasible.
So, we specify the following parametric forms for these functions.
We require $b$ to be a continuous function and so it is necessary that
$b(0)=0$.
Suppose that $x_1, \ldots, x_q$
satisfy $0 = x_1 < x_2 < \cdots < x_q = d$.
Obviously, $b(x_1)=0$, $b(x_q)=0$ and $s(x_q)=t_{n-p,1-\frac{\alpha}{2}}$.
The function $b$ is fully specified by the vector $\big (b(x_2), \ldots, b(x_{q-1}) \big)$
as follows. Because $b$ is assumed to be an odd function, we know that
$b(-x_i) = -b(x_i)$ for $i=2,\ldots,q$.
We specify the value of $b(x)$ for any $x \in [-d,d]$
by cubic spline interpolation for these given function values.
We specify the function $s$ by the vector $\big (s(x_1), \ldots, s(x_{q-1}) \big)$
as follows.
The value of $s(x)$ for any $x \in [0,d]$ is specified
by cubic spline interpolation for these given function values. We call $x_1, x_2, \ldots x_q$ the knots.
We have taken these knots to be equally spaced.

To conclude, the new $1-\alpha$ confidence interval for $\theta$ that utilizes the uncertain prior information
that $\tau=0$ is obtained as follows. Theoretically, the performance of the new confidence interval will
improve as $d$ increases and the spacing between the knots $x_i$ decreases. However, the computation of this
confidence interval becomes numerically unstable if $d$ is too large and/or the number of knots is too large.
So, for each candidate value of the parameter $\lambda$, we carry out the following computational procedure
for judiciously-chosen sets of values
of $d$ and knots $x_i$.

\medskip

\noindent {\underbar{Computational Procedure}} \newline
Let $\boldsymbol{z} = \big(b(x_2), \ldots, b(x_{q-1}), s(x_1), \ldots, s(x_{q-1}) \big )$.
Define $f(\boldsymbol{z})$ to be the objective function \eqref{criterion_final},
thought of as a function of $\boldsymbol{z}$. Also define $\tilde c(\boldsymbol{z}; \gamma)$ to
be  $(1-\alpha) -$ \eqref{cov_prob}, thought of as a function of $\boldsymbol{z}$ for given $\gamma$.
Minimize $f(\boldsymbol{z})$ subject to the constraints that $s(x) \ge 0$ for all
$x \in [0,d]$ and the nonlinear coverage constraints that $\tilde c(\boldsymbol{z}; \gamma)
\le 0$ for all $\gamma \ge 0$.
Plot the coverage probability of $J(b,s)$, as a function of $\gamma \ge 0$.
Also plot $e^2(\gamma;s)$, the square of the scaled expected length,
as a function of $\gamma \ge 0$.

\bigskip

Based on these plots, and possibly on the strength of our prior information that $\tau=0$,
we choose appropriate values of $\lambda$, $d$ and  knots $x_i$. Extensive guidelines for this choice are
presented in Section 4 of Kabaila and Giri (2009c).
The confidence interval corresponding to this choice is the new
$1-\alpha$ confidence interval for $\theta$.
The focus of the present paper is how this Computational Procedure can be made feasible and
practicable.


\bigskip

\noindent {\bf 3. Implementation of the coverage probability constraints}

\medskip

At first sight, the continuum of nonlinear coverage constraints $\tilde c(\boldsymbol{z}; \gamma)
\le 0$ for all $\gamma \ge 0$ would seem to make the Computational Procedure infeasible.
However,
Restriction 3 implies that, for
any reasonable choice of  $\boldsymbol{z}$, $\tilde c(\boldsymbol{z}; \gamma) \rightarrow 0$
as $\gamma \rightarrow \infty$. Also, for any given value of $\boldsymbol{z}$, $\tilde c(\boldsymbol{z}; \gamma)$
is a smooth function of $\gamma \ge 0$. This suggests that this continuum of constraints
can be replaced in the computations by
the following finite set of constraints:
$\tilde c(\boldsymbol{z}; \gamma) \le 0$  for every $\gamma \in \{0, \Delta, 2 \Delta, \ldots, M \}$, where
$\Delta$ is a sufficiently small positive number and $M$ is sufficiently large.
It is easy to check numerically whether or not given values of $\Delta$ and $M$ are adequate.
If the graph of $c(\gamma;b,s, \rho)$ falls below $1-\alpha$ for some values of $\gamma \ge 0$ then
this choice is inadequate. On the other hand, if $c(\gamma;b,s, \rho) \ge 1-\alpha$ for all $\gamma \ge 0$ then
this choice is adequate. This numerical check corresponds to the following easily-proved result.

\medskip

\noindent {\bf Lemma 1.} \ {\sl Suppose that $\boldsymbol{z}^*$ minimizes $f(\boldsymbol{z})$ subject to the constraints that $s(x) \ge 0$ for all
$x \in [0,d]$ and the coverage constraints that $\tilde c(\boldsymbol{z}; \gamma)
\le 0$ for all $\gamma \ge 0$. Also, suppose that $\boldsymbol{z}^{\prime}$ minimizes $f(\boldsymbol{z})$ subject to the constraints that $s(x) \ge 0$ for all
$x \in [0,d]$ and the coverage constraints that $\tilde c(\boldsymbol{z}; \gamma)
\le 0$ for every $\gamma \in \{0, \Delta, 2 \Delta, \ldots, M \}$. If $\tilde c(\boldsymbol{z^{\prime}}; \gamma)
\le 0$ for all $\gamma \ge 0$ then $f(\boldsymbol{z}^{\prime}) = f(\boldsymbol{z}^*)$.}

\medskip

\noindent For the numerical example presented in Section 7, we chose $\Delta=0.5$ and
$M = 50$. That this choice is adequate is clear from the plot of $c(\gamma;b,s, \rho)$, as a function of $\gamma \ge 0$,
in the top panel of Figure 2.

The function $s$ needs to satisfy the continuum of constraints $s(x) \ge 0$ for all $x \in [0,d]$.
Similarly to the coverage probability constraints, these could be replaced by the following finite set of constraints:
$s(x) \ge 0$ for each $x \in \{0, \delta, 2 \delta, \ldots, d \}$ where $\delta$ is a sufficiently small positive number.
However, it was found that the constraints $s(x_i) \ge \frac{1}{4} \, t_{n-p, 1 - \frac{\alpha}{2}}$ for $i=1, \ldots, q-1$
were not too restrictive and, in practice, guaranteed that $s(x) \ge 0$ for all $x \in [0,d]$.

The constrained minimization problem is solved numerically using the MATLAB function {\tt fmincon}.
The starting value of $\boldsymbol{z}$ was chosen to correspond to the
 standard $1-\alpha$ confidence interval $I$. In other words, for this starting value,
$b(x_2)=0, \ldots, b(x_{q-1})=0, s(x_1)=t_{n-p, 1 - \frac{\alpha}{2}}, \ldots,
s(x_{q-1})=t_{n-p, 1 - \frac{\alpha}{2}}$.
The ``Medium-Scale Optimization''
option for this function is used. This option uses a Sequential Quadratic Programming (SQP) method described in detail in the
documentation for the Optimization toolbox.

\bigskip

\noindent {\bf 4. Bounds on the truncation errors}

\medskip

The double integrals in \eqref{cov_prob}, \eqref{comp_conv_e} and
the second term on the right-hand-side of
\eqref{criterion_final} are evaluated as follows.
These integrals are first truncated with respect to $w$, followed by numerical evaluation of the truncated
double integrals using the MATLAB function {\tt dblquad}. In this section, we derive bounds on the resulting
truncation errors. We use $c$ to denote the upper endpoint of the truncated integral with respect to $w$.

Define the truncation error
\begin{align*}
e_1 &= \int_0^{\infty} \int_{-d}^{d} \big( k(wx,w, \gamma, \rho) - k^{\dag}(wx,w, \gamma, \rho) \big)
\, \phi(wx-\gamma)\,  dx \, w \, f_W(w) \, dw  \\
&\phantom{123456}- \int_0^c \int_{-d}^{d} \big( k(wx,w, \gamma, \rho) - k^{\dag}(wx,w, \gamma, \rho) \big)
\, \phi(wx-\gamma)\,  dx \, w \, f_W(w) \, dw  \\
&=\int_c^{\infty} \int_{-d}^{d} \big( k(wx,w, \gamma, \rho) - k^{\dag}(wx,w, \gamma, \rho) \big)
\, \phi(wx-\gamma)\,  dx \, w \, f_W(w) \, dw.
\end{align*}
As proved in Appendix B, $|e_1| \le P(Q > mc^2)$, where $Q\sim \chi^2_m$.

To find bounds on the other truncation errors, we will use the following lemma, which is proved in Appendix C.

\medskip

\noindent {\bf Lemma 2.}
\begin{equation*}
\int_c^{\infty} w f_W(w) dw = \sqrt{\frac{2}{m}} \, \frac{\Gamma \big(\frac{m}{2} +
\frac{1}{2}  \big)}{\Gamma \big(\frac{m}{2}\big)} P(\tilde Q > mc^2), \ \text{\sl where } \tilde Q \sim \chi^2_{m+1}.
\end{equation*}
%

\smallskip

\noindent Define the truncation error
\begin{align*}
e_2 &=  \int^{\infty}_0 \int^{d}_{- d} \left (s(|x|) - t_{n-p,1-\frac{\alpha}{2}} \right )
\phi(w x -\gamma) \, dx \,  w^2 \,  f_W(w)  \, dw \\
&\phantom{123456}- \int^c_0 \int^{d}_{- d} \left (s(|x|) - t_{n-p,1-\frac{\alpha}{2}} \right )
\phi(w x -\gamma) \, dx \,  w^2 \,  f_W(w)  \, dw \\
&= \int^{\infty}_c \int^{d}_{- d} \left (s(|x|) - t_{n-p,1-\frac{\alpha}{2}} \right )
\phi(w x -\gamma) \, dx \,  w^2 \,  f_W(w)  \, dw.
\end{align*}
As proved in Appendix D, $|e_2|$ is bounded above by
\begin{equation*}
\text{max}_{y\geq 0} \left|s(y) -
t_{n-p,1-\frac{\alpha}{2}} \right| \sqrt{\frac{2}{m}} \, \frac{\Gamma(\frac{m}{2}+1)}{\Gamma(\frac{m}{2})}
P(\tilde Q > mc^2),
\end{equation*}
where $\tilde Q \sim \chi^2_{m+1}$.

Define the truncation error
\begin{align*}
e_3 &=
 \int^{\infty}_0 \int^{d}_{0} \left (s(x) - t_{n-p,1-\frac{\alpha}{2}} \right )
\phi(w x) \, dx \,  w^2 \,  f_W(w)  \, dw \\
&\phantom{123456}-\int^c_0 \int^{d}_{0} \left (s(x) - t_{n-p,1-\frac{\alpha}{2}} \right )
\phi(w x) \, dx \,  w^2 \,  f_W(w)  \, dw \\
&= \int^{\infty}_c \int^{d}_{0} \left (s(x) - t_{n-p,1-\frac{\alpha}{2}} \right )
\phi(w x) \, dx \,  w^2 \,  f_W(w)  \, dw
\end{align*}
As proved in Appendix E, $|e_3|$ is bounded above by
\begin{equation*}
\frac{\text{max}_{y\geq 0} \left|s(y) -
t_{n-p,1-\frac{\alpha}{2}} \right|}{2} \sqrt{\frac{2}{m}} \, \frac{\Gamma(\frac{m}{2}+1)}{\Gamma(\frac{m}{2})}
P(\tilde Q > mc^2),
\end{equation*}
where $\tilde Q \sim \chi^2_{m+1}$.

We may, very conservatively, assume that $\text{max}_{y\geq 0} \left|s(y) -
t_{n-p,1-\frac{\alpha}{2}} \right| \le 10$ at or near the solution to the constrained minimization problem.
For any given value of $m$, it is easy to compute the values of $c$ such that these upper bounds on the
magnitudes of the the truncation errors are equal to some small specified positive number.

\newpage


\noindent {\bf 5. Some practical advice}

\medskip

It was found that the computation of the coverage probability \eqref{cov_prob}, which entails the computation of
\begin{equation}
\label{trouble}
\int_0^c \int_{-d}^{d} \big( k(wx,w, \gamma, \rho) - k^{\dag}(wx,w, \gamma, \rho) \big)
\, \phi(wx-\gamma)\,  dx \, w \, f_W(w) \, dw
\end{equation}
using the MATLAB function {\tt dblquad}, was inaccurate for small $n-p$.
The reason for this was found numerically to be the following. For small $n-p$ the integrand of
\eqref{trouble} is non-zero only for $w$ very close to zero; elsewhere in the interval $[0,c]$
it is very close to zero. As a result, {\tt dblquad} may largely ``miss'' the non-zero values
of the integrand, leading to the inaccurate computation of \eqref{trouble}. Our pragmatic solution
to this problem is as follows. If $c \ge 3$ then we perform two numerical integrations using {\tt dblquad}.
The first numerical integration evaluates
\begin{equation*}
\int_0^2 \int_{-d}^{d} \big( k(wx,w, \gamma, \rho) - k^{\dag}(wx,w, \gamma, \rho) \big)
\, \phi(wx-\gamma)\,  dx \, w \, f_W(w) \, dw
\end{equation*}
and the second numerical integration evaluates
\begin{equation*}
\int_2^c \int_{-d}^{d} \big( k(wx,w, \gamma, \rho) - k^{\dag}(wx,w, \gamma, \rho) \big)
\, \phi(wx-\gamma)\,  dx \, w \, f_W(w) \, dw
\end{equation*}
These two evaluations are then added to obtain the computed value of \eqref{trouble}.

To help prevent the occasional instability in the computation of the solution to the constrained minimization
problem, the following bounds were applied: $-100 \le b(x_i) \le 100$ for $i=2, \ldots, q-1$ and
$s(x_i) \le 200$ for $i=1, \ldots, q-1$.
 In some cases, it was found that the spline defining the function $b$ had oscillations that were clearly spurious.
These oscillations disappeared when the endpoint constraints $b^{\prime}(-d) = 0$ and  $b^{\prime}(d) = 0$
were introduced. These endpoint constraints are now part of the computational method.

The computation of the solution to the constrained minimization problem can be quite delicate,
especially when $n-p$ is small. In some cases the computation may not converge to the solution to the constrained minimization
problem, as evidenced by poor coverage properties of the computed confidence interval and/or spurious oscillations in the values
of $b$ and $s$. In this case, the computed solution is used as the starting value for another computation of the solution
to the constrained minimization problem. This procedure usually leads to the successful computation of the solution to this
minimization problem.

\bigskip

\noindent {\bf 6. Numerical example}

\medskip

Kabaila and Giri (2009c) present an example of the
new $1-\alpha$ confidence interval that utilizes the uncertain prior
information, for the case that $\rho = -1/\sqrt{2}$, $n-p=76$ and $1-\alpha=0.95$.
In the present section we consider the more computationally challenging case that $\rho=0.4$, $n-p=1$ and $1-\alpha=0.95$.
For each candidate value of the parameter $\lambda$, we carried out the Computational Procedure
(described in Section 2) for judiciously-chosen sets of values
of $d$ and knots $x_i$. Using the guidelines for the choice of $\lambda$, $d$ and the knots $x_i$
presented in Section 4 of Kabaila and Giri (2009c), we chose $\lambda=0.2$, $d=30$  and
the equidistant knots $x_i$ at $0,(d/6),\ldots,d$. The resulting functions $b$ and $s$, which specify the
new 0.95 confidence interval for $\theta$ that utilizes the uncertain prior information, are plotted in Figure 1.
The performance of this confidence interval
is shown in Figure 2. The top panel of this figure shows that the coverage probability of this confidence
interval is 0.95 throughout the parameter space. The bottom panel of Figure 2 is a plot
of the square of the scaled expected length of this confidence interval as a function of
$\gamma$. When the prior information is correct (i.e. $\gamma=0$), we gain since the square of the scaled expected length
is substantially smaller than 1. The maximum value of the square of the scaled expected length is not too large.
The new 0.95 confidence interval for $\theta$ coincides with the standard $1-\alpha$ confidence interval
when the data strongly contradicts the prior information. This is reflected in Figure 2 by the fact that the
square of the scaled expected length approaches 1 as $\gamma \rightarrow \infty$.

\begin{figure}[h]
\label{Figure1} \centering
\includegraphics[scale=0.75]{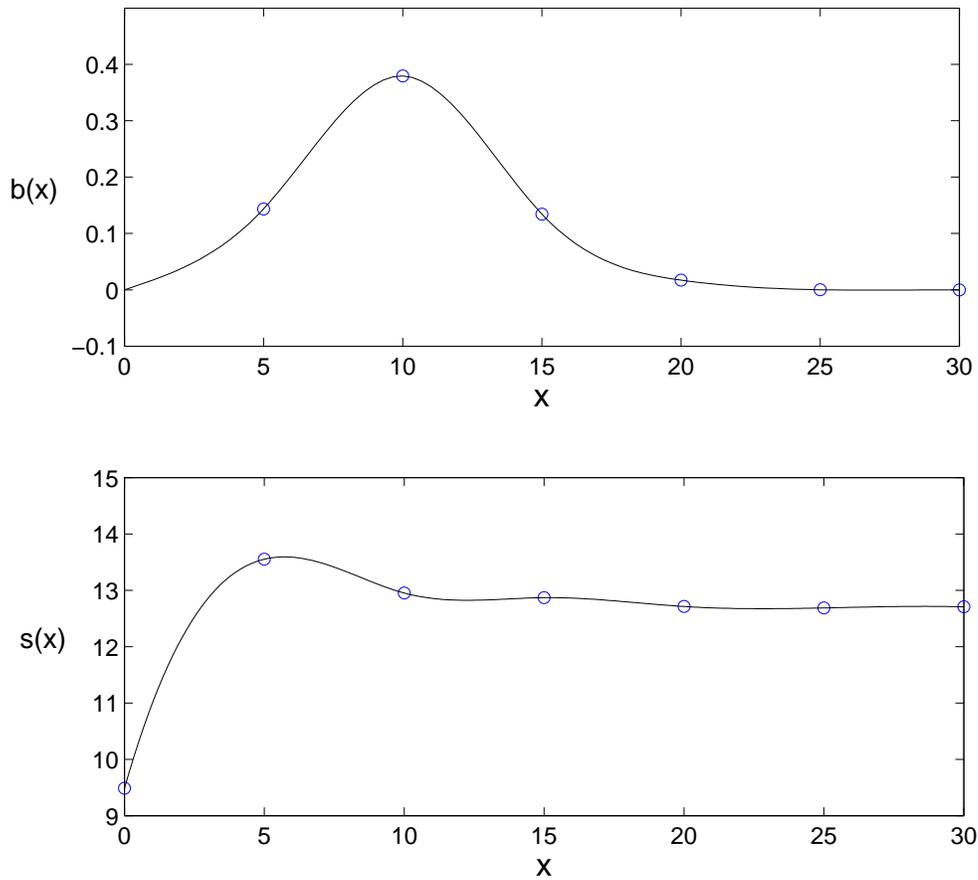}
       \caption{ Plots of the functions $b$ and $s$ for the new confidence interval for $\theta$
       when $\rho=0.4$, $n-p=1$ and $1-\alpha=0.95$.
       These functions are obtained using $d=30$, $\lambda=0.2$ and the equidistant knots $x_i$ at $0,(d/6),\ldots,d$.}
\end{figure}

\begin{figure}[h]
\label{Figure1} \centering
\includegraphics[scale=0.75]{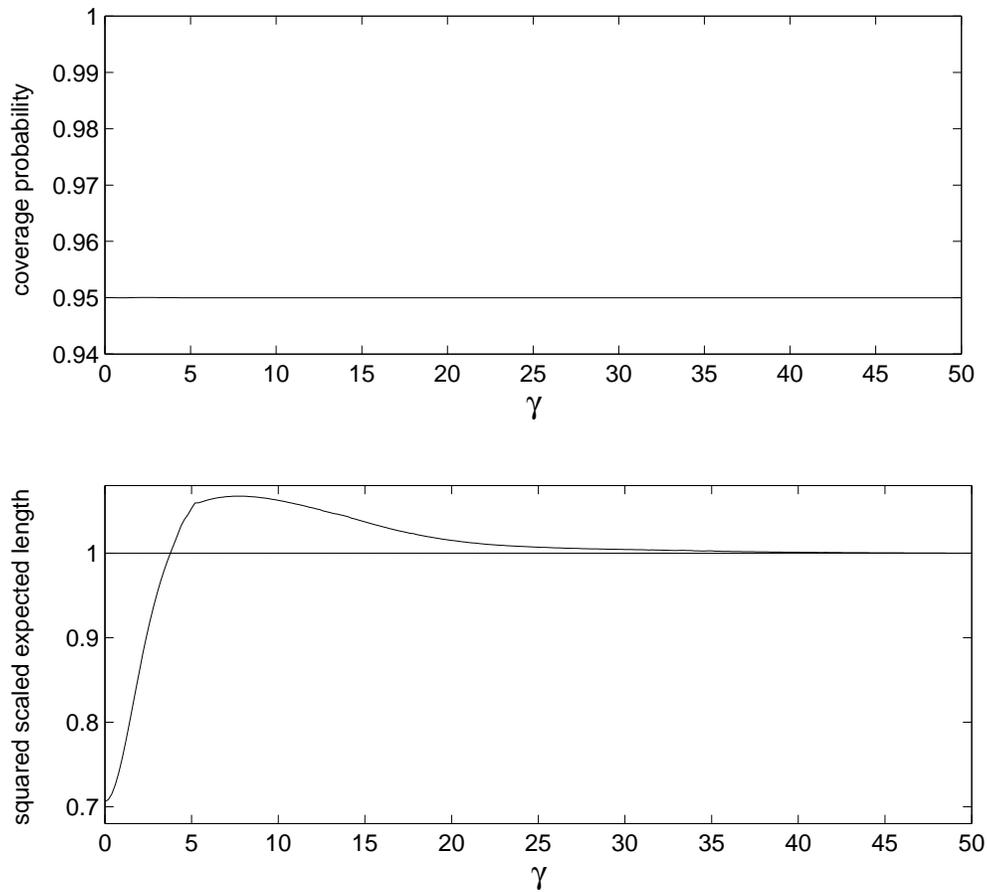}
       \caption{ Plots of the coverage probability and $e^2(\gamma;s)$, the squared scaled expected length,
as functions of $\gamma$  of the new 0.95 confidence interval for
$\theta$ when $n-p=1$, $\rho=0.4$.  These plots are obtained
using $d=30$, $\lambda=0.2$ and the equidistant knots $x_i$ at at
$0,(d/6),\ldots,d$.}
\end{figure}
\FloatBarrier

\bigskip

\noindent {\bf Appendix A. Computation of $\boldsymbol{E(W)}$}

\medskip

Using the well-known formula for the moments of a random variable with a gamma distribution
(see e.g. Casella and Berger (2002, p.130)),
it may be shown that
\begin{equation*}
E(W) = \sqrt{\frac{2}{m}} \, \frac{\Gamma \big(\frac{m}{2} +
\frac{1}{2}  \big)}{\Gamma \big(\frac{m}{2}\big)}.
\end{equation*}
By 6.1.47 on p.257 of Abramowitz and Stegun (1965), $E(W) \rightarrow 1$ as $m \rightarrow \infty$.
However, when $m$ is even moderately large, $\Gamma \big(m/2\big)$
is extremely large. To avoid problems with overflow, we first compute $\ln \big(\Gamma
\big(\frac{1}{2} + \frac{m}{2} \big) \big)$ and $\ln \big (\Gamma
\big(\frac{m}{2}\big) \big)$  by using the MATLAB function
{\tt gammaln}. We then find $E(W)$ by computing
\begin{equation*}
\exp \Big( \textstyle{ - \frac{1}{2} \ln \big(\frac{m}{2}\big) + \ln \big(\Gamma \big(\frac{1}{2} +
\frac{m}{2} \big) \big) - \ln \big (\Gamma \big(\frac{m}{2}\big)
\big) }\Big).
\end{equation*}

\bigskip

\noindent {\bf Appendix B. Derivation of the bounds on the truncation error $\boldsymbol{e_1}$}

\medskip

By changing the variable of integration from $x$ to $h=w x$ in the inner integral that defines $e_1$,
we obtain
\begin{equation*}
e_1 = \int_c^{\infty}
\int_{-dw}^{dw} \big( k(h,w, \gamma, \rho) - k^{\dag}(h,w, \gamma,
\rho) \big) \, \phi(h-\gamma)\,  dh \, f_W(w) \, dw.
\end{equation*}
It follows from the proof of Theorem 1 of Kabaila and Giri (2009c) that $k(h,w, \gamma, \rho)$ and
$k^{\dag}(h,w, \gamma,\rho)$ are conditional probabilities and so they belong to $[0,1]$.
Hence
\begin{equation*} -1\,\leq \int_{-dw}^{dw} \big( k(h,w, \gamma, \rho) - k^{\dag}(h,w,
\gamma, \rho) \big) \, \phi(h-\gamma)\,  dh \leq 1.
\end{equation*}
Therefore
\begin{equation*}
-\int_c^{\infty} \, f_W(w) \, dw \leq e_1 \leq\int_c^{\infty} \, f_W(w) \, dw.
\end{equation*}
Remember, $f_W(w) = 2 m w f_m(m w^2)$ for all $w > 0$, where
$f_m$ denotes the $\chi^2_m$ probability density function. Thus
\begin{equation*}
\int_c^{\infty} \, f_W(w) \, dw = \int_c^{\infty} \, 2mw \, f_m(m
w^2) \, dw.
\end{equation*}
Changing the variable of integration from $w$ to $y = mw^2$, we find that this integral is equal to
\begin{equation*}
 \int_{mc^2}^{\infty} \,  f_m(y) \, dy = P(Q > mc^2),
\end{equation*}
where $Q \sim \chi^2_m$. Hence $| e_1| \le P(Q > mc^2)$.

\bigskip

\noindent {\bf Appendix C. Proof of Lemma 2}

\medskip

Observe that
\begin{equation*}
\int_c^{\infty} w f_W(w) dw = \int_c^{\infty} w \, 2 m w \, f_m(mw^2) \, dw,
\end{equation*}
where, as in Section 2, $f_m$ denotes the $\chi_m^2$ probability density function.
Changing the variable of integration to $y = m w^2$, the right-hand-side becomes
\begin{align*}
&\sqrt{\frac{2}{m}} \, \frac{\Gamma \big(\frac{m}{2} + \frac{1}{2}  \big)}
{\Gamma \big(\frac{m}{2}\big)}
\int_{mc^2}^{\infty} \frac{1}{2^{(m+1)/2} \Gamma \big(\frac{m}{2} +
\frac{1}{2}  \big)} \, e^{-y/2} \, y^{((m+1)/2) - 1)} \, dy.
\end{align*}
The result follows from the fact that the integral in this expression is equal to
$P(\tilde Q > mc^2)$, where $\tilde Q \sim \chi^2_{m+1}$.

\bigskip

\noindent {\bf Appendix D. Derivation of the bounds on the truncation error $\boldsymbol{e_2}$}

\medskip

By changing the variable of integration from $x$ to $h=w x$ in the inner integral that defines $e_2$,
we obtain
\begin{equation*}
e_2
= \int^{\infty}_c \int^{d w}_{- d w} \left (s \left (\frac{|h|}{w} \right) - t_{n-p,1-\frac{\alpha}{2}} \right )
\phi(h -\gamma) \, dx \,  w \,  f_W(w)  \, dw.
\end{equation*}
Thus,
\begin{align*}
|e_2|
&\le \int^{\infty}_c \int^{d w}_{- d w} \max_{y \ge 0}\left |s \left (y \right) - t_{n-p,1-\frac{\alpha}{2}} \right |
\phi(h -\gamma) \, dx \,  w \,  f_W(w)  \, dw \\
&= \max_{y \ge 0}\left |s \left (y \right) - t_{n-p,1-\frac{\alpha}{2}} \right |
\int^{\infty}_c \int^{d w}_{- d w}
\phi(h -\gamma) \, dx \,  w \,  f_W(w)  \, dw \\
&\le \max_{y \ge 0}\left |s \left (y \right) - t_{n-p,1-\frac{\alpha}{2}} \right |
\int^{\infty}_c  \,  w \,  f_W(w)  \, dw.
\end{align*}
The result now follows from Lemma 2.

\bigskip

\noindent {\bf Appendix E. Derivation of the bounds on the truncation error $\boldsymbol{e_3}$}

\medskip

By changing the variable of integration from $x$ to $h=w x$ in the inner integral that defines $e_3$,
we obtain
\begin{align*}
|e_3| &\le \int^{\infty}_c \int^{dw}_{0} \left |s(h/w) - t_{n-p,1-\frac{\alpha}{2}} \right |
\phi(h) \, dh \,  w \,  f_W(w)  \, dw \\
&\le \int^{\infty}_c \int^{dw}_{0}  \max_{y \ge 0} \left |s(y) - t_{n-p,1-\frac{\alpha}{2}} \right |
\phi(h) \, dh \,  w \,  f_W(w)  \, dw \\
&\le \frac{\max_{y \ge 0} \left |s(y) - t_{n-p,1-\frac{\alpha}{2}} \right |}{2}
\int^{\infty}_c  w \,  f_W(w)  \, dw.
\end{align*}
The result now follows from Lemma 2.

\baselineskip=21pt

\bigskip

\noindent {\bf References}

\medskip

\rf Abramowitz, M., Stegun, I.A., 1965. Handbook of Mathematical Functions with Formulas,
Graphs, and Mathematical Tables. Dover, New York.

\smallskip

\rf Casella, G., Berger, R. L., 2002. Statistical
Inference, 2nd ed. Duxbury, Pacific Grove, California.

\smallskip

\rf Farchione, D., Kabaila, P., 2008. Confidence intervals for the normal mean utilizing prior information.
Statistics \& Probability Letters 78, 1094--1100.

\smallskip

\rf Giri, K., 2008. Confidence intervals in regression utilizing prior information.
Unpublished PhD thesis, August 2008, Department of Mathematics and Statistics, La Trobe University.

\smallskip

\rf Giri, K., Kabaila, P., 2008. The coverage probability of confidence intervals in $2^r$ factorial
experiments after preliminary hypothesis testing.
Australian \& New
Zealand Journal of Statistics 50, 69--79.

\smallskip

\rf Kabaila, P., 1995. The effect of model selection on confidence regions and
prediction regions. Econometric Theory 11,
537--549.

\smallskip

\rf Kabaila, P., 1998. Valid confidence intervals in
regression after variable selection. Econometric Theory 14,
463--482.

\smallskip

\rf Kabaila, P., 2005. On the coverage probability of confidence intervals
in regression after variable selection. Australian \& New Zealand Journal of Statistics
47, 549--562.

\smallskip

\rf Kabaila, P., 2009. The coverage properties of confidence regions after model selection.
To appear in International Statistical Review.

\smallskip

\rf Kabaila, P., Giri, K., 2009a. Large-sample confidence intervals for the treatment difference
in a two-period crossover trial, utilizing prior information. Statistics \& Probability
Letters 79, 652--658.

\smallskip

\rf Kabaila, P., Giri, K., 2009b. Upper bounds on the minimum coverage probability of confidence
intervals in regression after variable selection. To appear in Australian \& New Zealand Journal of Statistics. arXiv:0711.0993

\smallskip

\rf Kabaila, P., Giri, K., 2009c. Confidence intervals in regression utilizing prior information.
To appear in Journal of Statistical Planning and Inference. doi:10.1016/ \newline   j.jspi.2009.03.018

\smallskip

\rf Kabaila, P., Leeb, H., 2006. On the large-sample minimum coverage
probability of confidence intervals after
model selection. Journal of the American Statistical Association 101, 619--629.

\end{document}